\documentclass[11pt]{article}

\usepackage[numbers]{natbib}
\usepackage{url}
\usepackage{amsmath}
\usepackage{graphicx}
\usepackage{cancel}
\usepackage{nameref}

\begin{document}

\markboth{Keh-Fei Liu}{Lattice_nEDM}

\begin{center}
{\bf{\Large Lattice QCD and the Neutron Electric Dipole Moment}} 

\bigskip
{\bf  Keh-Fei Liu}

\end{center}

\begin{center}
{
Department of Physics and Astronomy, University of Kentucky, Lexington, KY 40506 \\
Nuclear Science Division, Lawrence Berkeley National Lab., Berkeley, CA 94720
} \\
\vspace{0.5cm}
{E-mail: liu@g.uky.edu}
\end{center}


\bigskip
\begin{abstract}
The recent lattice QCD calculations of the neutron and proton electric dipole moments (EDMs) and the CP-violating $\pi {\rm NN}$ coupling constant due to the $\theta$ term are reviewed. Progress towards nucleon EDM calculations, including the Weinberg three-gluon operator, and the quark chromoelectric dipole moment operator and their renormalization, is also discussed.
\end{abstract}



\bigskip
\tableofcontents

\section{INTRODUCTION}
The amount of nuclear matter we observe in the Universe must be the remnant of the imbalance between matter and antimatter during the evolution shortly after the Big Bang, commonly referred to as the baryonic asymmetry of the Universe (BAU). Observations indicate that the density of antimatter is much smaller than that of matter. For example, the antiproton to proton flux ratio is measured to be $\sim 10^{-4}$ by the AMS collaboration~\cite{AMS:2016oqu}. This asymmetry is also quantified by the ratio $\eta$, which is the net baryon density relative to the photon density: $\eta = \frac{n_b}{n_{\gamma}}$, where $n_b = n_B - n_{\overline{B}}$ is the difference between the baryon density $n_B$ and  the antibaryon density $n_{\overline{B}}$, and $n_{\gamma}$ is the photon density. This ratio can be determined in two different ways: one from the abundance of light elements in the intergalactic medium (IGM) ~\cite{Fields:2019pfx}, produced during the early universe in the process known as Big Bang nucleosynthesis (BBN) and the other from the power spectrum of temperature fluctuations in the cosmic microwave background (CMB)~\cite{Jungman:1995bz}; these methods yield consistent results. The latest value from BBN is $\eta \approx 6 \times 10^{-10}$~\cite{Yeh:2022heq,ParticleDataGroup:2024cfk}. Such a small baryon asymmetry poses one of the most outstanding problems facing cosmology, astrophysics and particle physics.  

According to Sakharov~\cite{Sakharov:1967dj}, there are three necessary conditions for baryogenesis 
: I) baryon number violation, II) charge (C) and charge-parity (CP) violation, and III) a deviation from thermal equilibrium. These conditions can be met in the standard model (SM). Conditions II) and III) are more straightforward. The weak interaction violates P-invariance maximally,
while CP-invariance is violated by the complex phase $\delta_{CKM}$ in the Cabibbo-Kobayashi-Maskawa (CKM)~\cite{Kobayashi:1973fv}  quark flavor-mixing matrix. The expansion of the Universe brings the primordial plasma out of thermal equilibrium. On the other hand, condition I) can occur due to the quantum anomaly which breaks baryon number conservation~\cite{Adler:1969gk,Bell:1969ts}. However, the strength of CP violation (CPV) in the CKM matrix is insufficient to explain  baryogenesis~\cite{Shaposhnikov:1987tw,Farrar:1993hn,Gavela:1993ts,Huet:1994jb}. Similarly, CP violation due to the QCD $\theta$ term is also unlikely to produce the observed baryon asymmetry in the Universe (BAU)~\cite{Dolgov:1991fr}. These considerations have led to the study of beyond standard model (BSM) physic scenarios for BAU~\cite{Cohen:1993nk,Chupp:2017rkp}. 

It is feasible to examine the Sakarov condition II) experimentally at low energies and theoretically in terms of the electric dipole moment (EDM). The discovery of a nonzero value of the permanent electric dipole moment (EDM) of non-degenerate systems, such as elementary particles, atoms, or molecules would signal the existence of both time-reversal (T) and parity (P) non-invariance, which is equivalent to CP violation~\cite{Luders:1954zz}. The earliest experiment to search for the permanent neutron electric dipole moment (nEDM)
aimed to test P non-invariance~\cite{Purcell:1950zz}. It was later realized that these experiments would also probe CP violation~\cite{Landau:1957tp,Lee:1973iz,Lee:1974jb}. 

Neutrons provide a convenient testing ground for EDM experimentally. Since neutrons do not carry electric charge, they can be subjected to large electric fields, either in a beam or in a bottle, without having their motion significantly affected. As spin 1/2 particles, neutrons can be readily spin-polarized using polarizing mirrors (in neutron beam) or superconducting magnets via the $\vec{\mu} \cdot \vec{B}$ interaction (for ultracold neutrons (UCN)). Searches for nEDM have been conducted in both beam and ultracold experiments with polarized neutrons. The latest measured value of nEDM from the ultracold neutron experiment is $d_n = (0.0 \pm 1.1_{\rm stat} \pm 0.2_{\rm sys}) \times 10^{-26}\, {\rm e\cdot cm}$ ~\cite{Abel:2020gbr}. Several ongoing experimental efforts include nEDM\@LANL~\cite{Ito:2017ywc}, TUCAN~\cite{TUCAN:2018vmr}, PanEDM~\cite{Wurm:2019yfj}, PNPI nEDM~\cite{Serebrov:2017rby} (these are UCN experiments), as well as beam nEDM~\cite{Chang:2018uxx} (a pulsed neutron beam experiment). While they have different timelines, they all aim to increase statistics by a factor 10 to 100 by 2030. 

The electric dipole moment (EDM) of charged particles can be measured in storage rings. For example, the measurement of the electron EDM (eEDM) has been attempted using trapped molecular ions~\cite{Cairncross:2017fip}, and there is a proposal to probe it
in a storage ring~\cite{Suleiman:2021whz}. The Storage Ring COSY plans to hold protons and deuterons for their EDM measurements~\cite{CPEDM:2019nwp}. Atoms and molecules have served as sensitive platforms for precision measurements of symmetry violations for decades, including CP violation for EDMs~\cite{Chupp:2017rkp,Safronova:2017xyt}. To investigate  hadronic CPV, diamagnetic systems are used, such as $^{129}{\rm Xe}, ^{199}{\rm Hg}$ and $^{225}{\rm Ra}$. These experiments currently set the best limits on the eEDM, semileptonic CPV interactions, and quark chromo-EDMs. They are also competitive with nEDM experiments regarding sensitivity to quark EDMs and the QCD $\theta$ term. A summary of the recent experiments on EDMs of various systems and their bounds is provided in Ref.~\cite{Chupp:2017rkp}.

There are two sources of CP violation in the standard model that contribute to the neutron electric dipole moment (nEDM). One is the CKM matrix in the weak interaction, which induces an EDM of the quarks at the two-loop order~\cite{Shabalin:1978rs,Shabalin:1982sg} involving heavy quarks. The recent results from the CKM matrix calculation of nEDM 
$\rm d_n$ and proton EDM (pEDM) $\rm{d_p}$
are \mbox{$1 \times 10^{-32}\, {\rm e \cdot cm} < \{ \rm{|d_n (CKM)|, |d_p (CKM)|}\}  < 6 \times 10^{-32}\, {\rm e \cdot cm}$}~\cite{Seng:2014lea}. This range is 5 to 6 orders of magnitude smaller than the current upper limit on the nEDM. Thus, nEDM experiments have a good discovery potential for BSM physics. 

The second source is the $\theta$-term in QCD, represented by the topological charge operator 
$-\bar{\theta} \frac{g^2}{32\pi^2} G_{\mu\nu}^a \tilde{G}_{\alpha\beta}^a$. The EDM induced by the CKM matrix is suppressed due to the flavor non-diagonal nature of quark flavor-mixing. The topological charge term is the only four-dimensional flavor-diagonal operator that can produce an EDM. If this is the sole contribution to the EDM, the current experimental limit on EDM implies that 
$\bar{\theta} < 10^{-10}$ based on chiral symmetry considerations of a pion loop contribution~\cite{Crewther:1979pi,Baluni:1978rf}. In contrast to the CKM matrix where $\delta_{CKM}$ is of order unity, this small $\bar{\theta}$ is known as the strong CP problem. 

One proposed solution to address the strong CP problem is the introduction of the axion - a pseudo-Goldstone boson - via the Peccei-Quinn mechanism~\cite{Peccei:1977hh}. In this framework, $\bar{\theta}$ becomes a dynamical field with a potential minimized at 
$\bar{\theta} = 0.$  The smallness of $\bar{\theta}$ can be understood as being dynamically relaxed~\cite{Peccei:1977ur,Peccei:1977hh,Weinberg:1977ma,Wilczek:1977pj}, though generally not to exactly zero, due to effects that violate the Peccei-Quinn symmetry~\cite{Barr:1992qq,Kamionkowski:1992mf,Holman:1992us,Ghigna:1992iv}.
Another possibility is that there are BSM interactions responsible for CP violation, which can be classified as higher dimensional operators in effective theories. Some of these operators can mix with the four-dimensional topological term. The smallness of 
$\bar{\theta}$ would then reflect the large scale of BSM physics.

While BAU and nEDM may require new beyond Standard Model (BSM) physics, relating nuclear and atomic EDMs
to the high-energy CP violation mechanism  will necessitate hadronic and nuclear inputs. 
Below the BSM scale, flavor-diagonal CP violation can be described by 
extending the Standard Model (SM) Lagrangian with gauge-invariant higher-dimensional operators, leading to a standard model effective field theory (SMEFT)~\cite{Buchmuller:1985jz,Grzadkowski:2010es,Cirigliano:2013lpa}. A list of operators relevant to CP violation can be found in~\cite{Engel:2013lsa}.
To address hadron and nuclear systems below the electroweak scale, heavy SM degrees of freedom, such as W and Z 
bosons and the Higgs, can be integrated out by matching the SMEFT Lagrangian to a low-energy effective field theory (LEFT)~\cite{Jenkins:2017jig,Jenkins:2017dyc}. The complete one-loop matching calculation has been carried out~\cite{Dekens:2019ept}. A subset of operators relevant to EDM, estimated to have dominant contributions~\cite{deVries:2012ab}, include the quark EDM (qEDM), the quark chromo-EDM (qcEDM), the gluon chromo-EDM (gcEDM) --- known as the Weinberg operator~\cite{Weinberg:1989dx} --- and the four-fermion operators. 

The discovery of an electric dipole moment (EDM) in any single system --- be it a hadron, a lepton or a molecule --- would represent a significant breakthrough and paradigm-shift. However, it would not be sufficient to definitely identify the source or discriminate among different BSM models. It is necessary to bound EDMs in complementary systems to identify the characteristic features of CP violation at low-energies and establish a connection to the underlying BSM origin. One crucial ingredient in this process is having reliable theoretical estimates of the relevant matrix elements for the aforementioned  CP-violating operators. This is where lattice QCD calculations come into play. 
  
Over the last five decades, lattice QCD calculations based on the Euclidean path-integral formulation of QCD on a discrete space-time lattice~\cite{Wilson:1974sk} have developed into a powerful tool for {\it ab initio} calculations of strong-interaction physics. It is presently the only theoretical approach to solving QCD with controlled systematic errors. Many of the recent lattice ensembles have reached the physical pion mass, allowing for systematic errors in continuum and infinite volume extrapolations to be quantified, controlled, and improved.

Lattice fermion formulations that accommodate chiral symmetry have been developed 
over the last three decades, including domain-wall fermions (DWF)~\cite{Kaplan:1992bt,Shamir:1993zy}, 
overlap fermions~\cite{Neuberger:1997fp}, and fixed-point fermions~\cite{Hasenfratz:1998jp}. 
In this review, we will discussion the progress on lattice calculations of the $\theta$ term, the qEDM, the qcEDM, and the Weinberg operator. A previous review on the subject has been given in \cite{Shindler:2021bcx}. This review serves as an update on the latest lattice results, particularly concerning the  $\theta$ term and the CP-violating $\pi {\rm NN}$ coupling. We will not discuss the four-fermion operators as there are currently no lattice calculations for these operators.

\section{QCD $\theta$ Term}

QCD can naturally incorporate a CP-violating term known as the topological $\theta$ term, which is odd under P and T transformations. The QCD+ $\bar{\theta}$ Lagrangian is then
\begin{equation}  \label{theta-L}
    \mathcal{L}_{{\rm QCD} + \bar{\theta}} = - \frac{1}{4} G_{\mu\nu}^a G^{a \mu\nu} + \overline{\psi} (-i D\!\!\!\!/ - M)\psi - \bar{\theta}\frac{g^2}{32\pi^2} G_{\mu\nu}^a \tilde{G}^{a \mu\nu}
    \end{equation}
where $\tilde{G}^{a \mu\nu}\equiv \epsilon^{\mu\nu\alpha\beta} G_{\alpha\beta}^a$ and $M$ is the quark matrix.

In general, the quark mass can be complex. When performing a chiral transformation
to redefine the quark fields, $\psi \rightarrow e^{i\alpha \gamma_5} \psi$
and $\overline{\psi} \rightarrow \overline{\psi} e^{i\alpha \gamma_5}$, the fermion determinant in the path integral~\cite{Fujikawa:1979ay} induces an additional measure in the form of the topological charge. This is a reflection of the anomalous $U_A(1)$ Ward identity
\begin{equation}   \label{AWI}
    \partial^{\mu} A_{\mu}^0 = \sum_{f}^{N_f} 2m_f P_f + \frac{N_f g^2}{16\pi^2}G_{\mu\nu}^a \tilde{G}^{a \mu\nu}
\end{equation}
where $A_{\mu}^0$ is the flavor-singlet axial-vector current, $P_f$ is the quark pseudoscalar density of the $f$-flavor, and $N_f$ is the number of flavors. This indicates that the conservation of axial current is violated due to quantum effects. 

The additional topological term can be
absorbed into the original one through the substitution $\bar{\theta} = \theta + arg det(M)$. This makes the quark matrix real, and the Lagrangian in Eq.~(\ref{theta-L}) already reflects this choice. This is a common approach where the $\theta$ term only involves the gluon field tensor, i.e. 
$E\cdot B$. Conversely, starting from the Lagrangian in Eq.~(\ref{theta-L}) and using a $U_A(1)$ chiral transformation~\cite{Crewther:1979pi,Baluni:1978rf}
to rotate the $\theta$ term into the pseudoscalar mass term~\cite{Baluni:1978rf}
for $N_f =3$ results in:
\begin{equation}   \label{theta-mass}
    \mathcal{L}_{\theta} \longrightarrow i\bar{\theta}\, m^{*} (\bar{u}\gamma_5 u +\bar{d}\gamma_5  + \bar{s}\gamma_5 s),
\end{equation}
where $m^{*} = \frac{m_u m_d m_s}{m_u m_d+ m_u m_s + m_d m_s}$. This basis is more accessible for calculations in chiral effective theories. 

The QCD action for the Lagrangian in Eq.~(\ref{theta-L}) in Euclidean space is given by:
\begin{equation}
    S_{QCD+\bar{\theta}}^E = \int d^4 x\, \lbrack\frac{1}{4} G_{\mu\nu}^a G_{\mu\nu}^a + \bar{\psi}(D\!\!\!\!/ + M)\psi -i\bar{\theta} \frac{g^2}{32\pi^2} G_{\mu\nu}^a \tilde{G}_{\mu\nu}^a \rbrack.
\end{equation}
where $Q=\frac{g^2}{32\pi^2} G_{\mu\nu}^a \tilde{G}_{\mu\nu}^a$ is the topological charge. Thus, we have: 
\begin{equation}
S_{QCD+\bar{\theta}}^E = S_{QCD}^E - i \bar{\theta}Q. 
\end{equation}
Since the $\theta$ term is imaginary, the usual probability based Monte Carlo approach encounters a sign problem. However, considering that $\bar{\theta} \ll 1$, one can Taylor expand in $\bar{\theta}$ to calculate observables in the presence of $Q$:
\begin{eqnarray} \label{Taylor}
    \langle O(\psi, \bar{\psi}, U)\rangle_{\bar{\theta}} &=& \frac{ \int 
    \mathcal{D}\psi\, \mathcal{D}\bar{\psi}\,\mathcal{D}U\, O(\psi, \bar{\psi}, U)\, e^{-S_{QCD}^E - i \bar{\theta}Q}}{\int \mathcal{D}\psi\, \mathcal{D}\bar{\psi}\,\mathcal{D}U\, e^{-S_{QCD}^E - i \bar{\theta}Q}},  \nonumber \\
    &&\longrightarrow \langle O\rangle_{\bar{\theta}=0} - i \bar{\theta}\langle O\,Q\rangle_{\bar{\theta}=0}   + \mathcal{O} (\bar{\theta}^2) . 
\end{eqnarray}
Thus, to first order in $\bar{\theta}$, one can calculate $\langle O\,Q\rangle_{\bar{\theta}=0}$. Here, $U$ represents the gauge link variable in the lattice formulation, and there are various gauge and fermion actions used on the lattice. We shall use Euclidean notation in the following text. 

The common calculation of the neutron electric dipole moment (nEDM) is conducted through
the vector form factor in the presence of the $\theta$ term in the action. In this case, the vector current form factor is given by~\cite{Abramczyk:2017oxr}
\begin{equation}
    \langle N(p')|J_{\mu}^{em}|N{p}\rangle_{\bar{\theta}} = \bar{u}_{p'} \Gamma_{\mu} u_{p}, 
\end{equation}
where
\begin{equation}
    \Gamma_{\mu} = \gamma_{\mu} F_1(q^2) - \frac{\sigma_{\mu\nu}q_{\nu}}{2M_N} [F_2(q^2) - i F_3 (q^2) \gamma_5] + \frac{F_A(q^2)}{m_N} (q\!\!\!/ \,q_{\mu}-q^2\gamma_{\mu})\gamma_5.
    \end{equation}
Here, $\sigma_{\mu\nu} = [\gamma_{\mu}, \gamma_{\nu}]/2i$, $F_3(q^2)$ is the electric dipole form factor, and $F_A(q^2)$ is the anapole form factor. 

It has been noted~\cite{Shintani:2005xg,Liu:2008gr} that the linear $\bar{\theta}$ dependence comes in two places: one is the insertion of $-i\bar{\theta}\,Q$ in addition to the current $J_{\mu}^{em}$, and the other is through the nucleon state, which acquires a CP-breaking phase. Specifically, the nucleon spinor in the $\bar{\theta}$ vacuum becomes:
\begin{equation}   \label{alpha5}
    u_{\bar{\theta}} = e^{i \bar{\theta}\alpha_5\gamma_5}u, \hspace{1.5cm} \bar{u}_{\bar{\theta}} = \bar{u}e^{i \bar{\theta}\alpha_5\gamma_5}.
\end{equation}
These spinors satisfy the following Dirac equations:
\begin{equation}
    [-ip\!\!\!/ - M_N^{\bar{\theta}}e^{-i\bar{\theta} \alpha_5 \gamma_5}]u_{\bar{\theta}} = 0 =
    \bar{u}_{\bar{\theta}}[-ip\!\!\!/ - M_N^{\bar{\theta}}e^{-i \bar{\theta} \alpha_5 \gamma_5}],
\end{equation}  
where $M_N^{\bar{\theta}} = M_N + \mathcal{O}(\bar{\theta}^2)$, so that 
\begin{equation}
    u_{\bar{\theta}}(p)\, \bar{u}_{\bar{\theta}}(p) = \frac{-i p\!\!\!/ 
    - M_N e^{2i\bar{\theta} \alpha_5 \gamma_5}}{2M_N}.
\end{equation}
It was shown~\cite{Abramczyk:2017oxr} that the combinations 
\begin{equation} \label{F3F2}
    F_3(q^2) = \tilde{F}_3(q^2) + 2\alpha_5 \tilde{F}_2 (q^2), \hspace{1.5cm} F_2(q^2) = \tilde{F}_2(q^2) - 2\alpha_5 \tilde{F}_3 (q^2)
\end{equation}
transform as an axial vector and vector, respectively. 
 Thus, the electric dipole moment is given by:
\begin{equation}
    d_N = \lim_{q^2 \rightarrow 0} \frac{F_3(q^2)}{2M_N}
\end{equation}
where $\tilde{F}_3$ and $\tilde{F}_2$ are those calculated
without considering the CP phase in the nucleon state. 
Since the electric dipole form factor will be determined from the ratio of the three-point to two-point correlators, the above results in Eq.~(\ref{F3F2}) can also be obtained from the $Q$ insertion separately in the three- and two-point functions without considering the CP phase in the nucleon state, following the expansion in Eq.~(\ref{Taylor}). Previous lattice nEDM calculations~\cite{Shintani:2005xg,Berruto:2005hg,Guo:2015tla,Shintani:2015vsx,
Alexandrou:2015spa} did not take the $2\alpha \tilde{F}_2$ term into account. When it was included, the results were within one sigma of zero, indicating that no nEDM had been observed~\cite{Abramczyk:2017oxr}. 

Furthermore, directly calculating with re-weighting of the total topological charge $Q$ in the evaluation of the correlator $\langle \mathcal{O}Q\rangle$ in Eq.~(\ref{Taylor}) is not practical. The average $|Q|$ in an ensemble grows with $\sqrt{V}$, the square root of the volume, which induces large fluctuations and negatively affects the signal-to-noise ratio (SNR) as $V$ increases. It has been found that considering  the total $Q$ as the sum of the local charge $q(x)$ ( i.e., $Q=\int d^4 x \,q(x)$) and summing a subset of
$q$ over limited time slices straddling the current time position reduces the error~\cite{Shintani:2015vsx}.
This is a consequence of cluster decomposition~\cite{Liu:2017man}, and one can use this property to reduce variance through a cluster decomposition error reduction technique (CDER)~\cite{Liu:2017man}, which we will discuss later.

New lattice efforts have been initiated~\cite{Syritsyn:2019vvt,Dragos:2019oxn,Alexandrou:2020mds,Bhattacharya:2021lol,Liang:2023jfj,He:2023gwp}. 
One lattice calculation~\cite{Dragos:2019oxn} used the clover fermions on a lattice with a spacing of $a = 0.0907\, {\rm fm}$ and investigated three pion masses at 410, 568, and 699 MeV for chiral extrapolation. Additionally, it included three lattice spacings of 0.0684, 0.0936 and 0.1095 fm, with a pion mass around 680 MeV for continuum extrapolation. The authors employed gradient flow~\cite{Luscher:2010iy,Luscher:2013cpa} to define the local topological charge, which yields integer $Q$ and does not require renormalization. They also adopted
a cluster decomposition error reduction technique (CDER) for the time slices to reduce errors in both the two- and three-point functions. The results obtained were $d_n= - 0.00152(71)\, \bar{\theta}\, e \cdot {\rm fm}$ and $d_p= 0.0011(10)\, \bar{\theta}\, e \cdot {\rm fm}$. 
The reported nucleon Schiff moments reported are $S_n = - 0.10(43) \times 10^{-4} \,\bar{\theta}\, e \cdot {\rm fm}$ and $S_p = 0.50(59) \times 10^{-4}\, \bar{\theta}\, e \cdot {\rm fm}$. 

Another calculation using $N_f = 2+1+1$ twisted mass and clover-improved fermion action was conducted by the Extended Twisted Mass Collaboration (ETMC)~\cite{Alexandrou:2020mds}. The lattice has a physical size of 5.13 fm with a spacing of approximately $a \simeq 0.08$ fm, and a pion mass at the physical value of 139 MeV. The fermion version (i.e., pseudoscalar quark bilinear operator) derived from the space-time integral of the anomalous Ward identity in Eq.~(\ref{AWI}) (where the divergence of the axial vector current drops out) was used for the topological charge operator $Q$. This formulation was found to be less noisy than that constructed from the gauge link variables, whether using gradient flow, link smearing, or cooling~\cite{Alexandrou:2017bzk}, and has smaller discretization errors. They obtained the nEDM value as \mbox{$|d_n| = 0.0009(24)\, \bar{\theta}\, e \cdot {\rm fm}$.} 

The \mbox{Los Alamos} group utilized the valence clover action on the $N_f = 2+1+1$ highly improved staggered quark (HISQ) gauge configurations for their nEDM calculation. They worked with nine ensembles featuring lattice spacings
ranging from 0.0570 fm to 0.1207 fm and pion masses from the physical value of 135 MeV to 320 MeV~\cite{Bhattacharya:2021lol}. The topological charge operator was obtained from the gradient flow
and the $\mathcal{O}(a)$ effects were considered due to the non-chiral valence quarks~\cite{Chen:2007ug} and
the fact that the vector current is not $\mathcal{O}(a)$ improved. Their chiral extrapolation fits, which included chiral logarithms, yielded $d_n = - 0.003(7)(20)\, \bar{\theta}\, e \cdot {\rm fm}$ and \mbox{$d_p = 0.024(10)(30)\, \bar{\theta}\, e \cdot {\rm fm}$.} 

\begin{figure}[htb]
\centering
\includegraphics[width=1.0\textwidth]{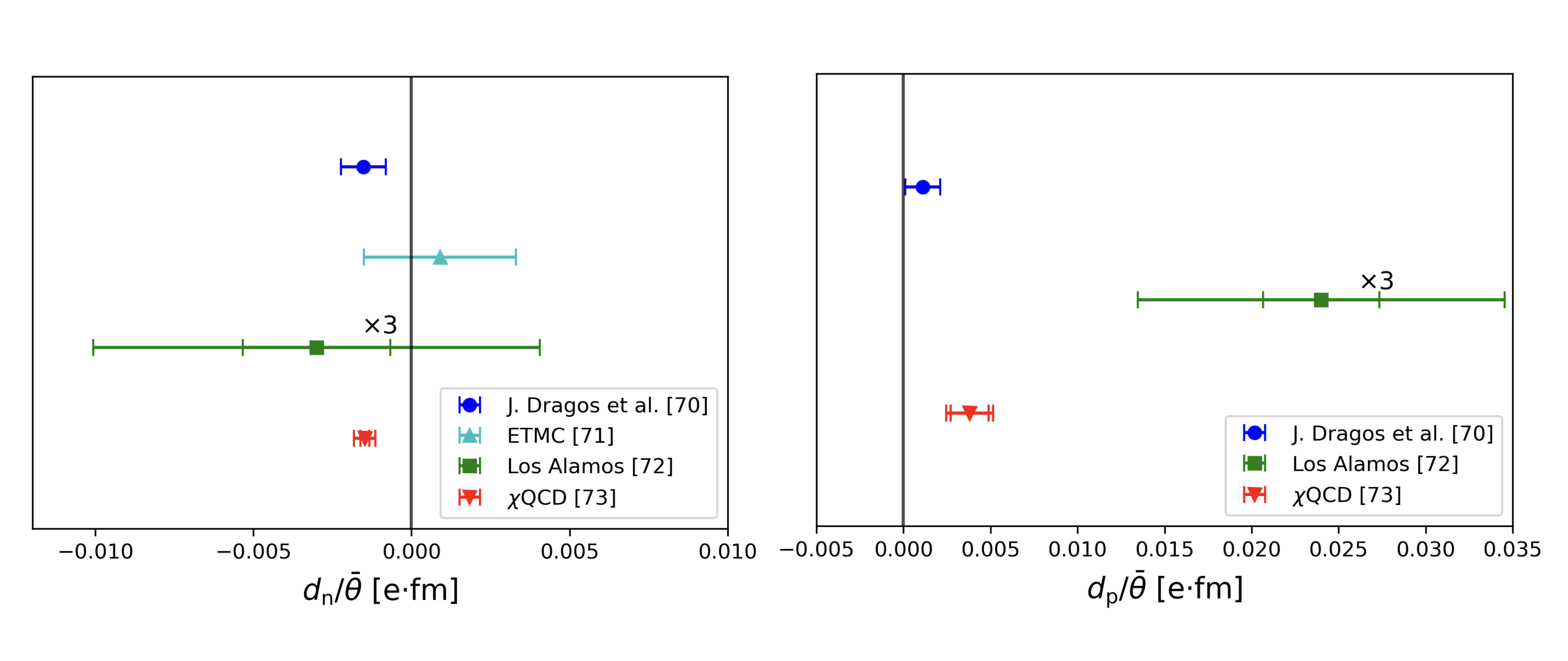}
\vspace{-0.5cm}
\caption{The lattice calculation results in Table~\ref{tab1} are plotted. The left and right panel is for nEDM and pEDM respectively. The legends give their reference numbers.}
\label{fig1}
\end{figure}

A recent calculation on the subject was carried out by the $\chi$QCD collaboration~\cite{Liang:2023jfj}. They employed overlap fermions on $2+1$-flavor domain-wall fermion (DWF) configurations on $24^3 \times 64$ lattices for three ensembles with sea pion masses
at 339, 432, and 560 MeV. For each ensemble with a different sea quark mass, they calculated 3 to 4 valence quark masses for partially quenched analyses. The local topological charge operator was defined with the overlap operator for consistency~\cite{Fujikawa:1998if,Adams:1998eg}. Thanks to the chiral nature of both the valence overlap fermions and the sea DWF, they achieved results with smaller
statistical errors: $d_n = - 0.00148 (14) (31)\, \bar{\theta}\, e \cdot {\rm fm}$ and $d_p = 0.0038 (11) (8)\, \bar{\theta}\, e \cdot {\rm fm}$. We shall present some details of this calculation in the following. 
The results of these calculations are tabulated in Table~\ref{tab1} and plotted in Fig.~\ref{fig1}.

\begin{table}[h]  
\tabcolsep7.5pt
\caption{Summary of recent lattice calculations of the neutron electric dipole moment (nEDM) and proton electric dipole moment (pEDM) from the $\theta$ term, including the quark actions, lattice spacings, and pion masses.}
\label{tab1}
\begin{center}
\begin{tabular}{@{}l|c|c|c@{}}
\hline
Action (ref.)  & a (fm)/ $m_{\pi}$ (MeV)& $d_n\, (\bar{\theta}\, e\cdot {\rm fm})$ & $d_p\,(\bar{\theta}\, e\cdot {\rm fm})$\\
\hline
Clover~\cite{Dragos:2019oxn} &0.0907 / 410, 568, 699 & -0.00152(71) & 0.0011(10)\\
     &0.0684, 0.0936, 0.1995 / $\sim 680$ & & \\
    \hline
Twisted Mass~\cite{Alexandrou:2020mds} &0.08 / 139 &0.0009(24) & - \\
\hline
Clover+HISQ~\cite{Bhattacharya:2021lol}& 0.0570 - 0.1207 / 135 - 320& - 0.003(7)(20) & 
0.024(10)(30) \\
\hline
Overlap+DWF~\cite{Liang:2023jfj}& 0.0114 / 339, 432, 560 & - 0.00148 (14) (31) & 0.0038 (11) (8) \\
\hline
\end{tabular}
\end{center}
\end{table}

Other efforts calculating the nEDM from the $\theta$ term using Domain-Wall-Fermions (DWF) has also been attempted~\cite{Syritsyn:2019vvt,He:2023gwp}. A background field approach~\cite{He:2023gwp} with DWF has reported preliminary results for nEDM at two pion masses: $d_n /\bar{\theta}\, (340 {\rm MeV}) = - 0.0072(20), \,\, $and$ \,\, d_n /\bar{\theta}\, (420 {\rm MeV}) = -0.0060 (20)$.  

\subsection{Overlap fermions}

It is shown~\cite{Luscher:1998pqa} that the lattice fermion action is invariant under the chiral transformation \\
\mbox{$\psi \rightarrow e^{i\alpha \gamma_5 (1 - 1/2 aD)}\,\psi$} and $\bar{\psi} \rightarrow 
\bar{\psi}\, e^{i\alpha (1 - 1/2 aD)\gamma_5}$, if the Dirac operator $D$ satisfies the Ginsparg-Wilson (GW) relation $\{\gamma_5, D\} = a\, D\gamma_5 D$~\cite{Ginsparg:1981bj}. At the continuum limit, the usual chiral anti-commutation relation $\{\gamma_5, D\}= 0$ is recovered. 

The overlap fermion formulation~\cite{Neuberger:1997fp} provides a robust framework for chiral symmetry on the lattice and defines a continuum like quark propagator and the local topological charge.
\begin{itemize}
\item {\bf GW Relation}: Overlap fermions satisfy
the Ginsparg-Wilson relation exactly, while the DWF~\cite{Kaplan:1992bt,Shamir:1993zy} approximate it, introducing a small residual mass due to the finite fifth dimension $L_s$.   This means overlap fermions can effectively retain chiral symmetry even at finite lattice spacing.

\item {\bf Massless Dirac Operator}: The massless overlap Dirac operator is given by:
\begin{equation}
D_{ov} = 1 + \gamma_5 \epsilon(H_w),
\end{equation} 
where $\epsilon(H)$ is a matrix sign function $\epsilon (H_w) = \frac{H_w(\rho)} {\sqrt{H_w(\rho)^2}}$ of the Hermitian Wilson-Dirac operator $H_w(\rho)= \gamma_5 D_w(\rho)$  with a negative mass parameter $\rho$. When the sign function is calculated with low eigenmodes projected from $H_w(\rho)$, the GW-relation can be satisfied to machine precision~\cite{Edwards:1998yw,Dong:2000mr,xQCD:2010pnl}. 

\item{\bf Effective Propagator}: The effective quark propagator for the massive overlap fermion takes the form~\cite{Chiu:1998gp,Liu:2002qu}
\begin{equation}
 D_{\rm eff}(m)^{-1} = (1 - \frac{D_{ov}}{2})D_{ov}(m)^{-1} 
= (D_c +m )^{-1}, 
\end{equation}
where $D_c = \rho D_{ov}/(1 - \frac{D_{ov}}{2})$ satisfies the anti-commutation relation $\{\gamma_5, D_c\} = 0$. This ensures that the spectrum behaves similarly to continuum fermions, with eigenvalues located on the imaginary axis. Whereas, the eigenvalues of $D_{ov}$ are on a circle in the complex plan~\cite{Liu:2002qu}. 

\item {\bf Topological Charge and Index Theorem}:
The Atiyah-Singer index theorem holds for overlap fermions, i.e., $n_{-} - n_{+} = Q$ where $n_{-}/n_{+}$ is the number of the left-handed/right-handed zera modes. $Q$ is the topological charge which
is derived from the overlap operator~\cite{Hasenfratz:1998ri,Luscher:1998pqa} 
\begin{equation}
  Q =  \frac{1}{2\rho} {\rm Tr}(\gamma_5 D_{ov})
\end{equation}
for each gauge configuration.

\item {\bf Anomalous Ward Identity}:
Furthermore, the anomalous Ward identity in Eq.~(\ref{AWI}) is satisfied~\cite{Hasenfratz:2002rp} with the local topological charge~\cite{Fujikawa:1998if,Adams:1998eg} 
\begin{equation} \label{local_q}
    q(x) =   \frac{1}{2\rho} tr_{cs} (\gamma_5 D_{ov}(x,x)),
\end{equation}
where the trace is over the color and spin indices, and the pseudoscalar density is \\
 \mbox{$P = \bar{\psi} \gamma_5 (1 - \frac{1}{2\rho}D_{ov})\psi$}. The axial current is derived through introducing a local $U(1)$ phase, \\
 \mbox{$A_{\mu}^0 = \frac{1}{2} \bar{\psi} (-\gamma_5 K_{\mu}(x) +K_{\mu}(x) \gamma_5(1 - \frac{1}{2\rho}D_{ov}))\psi$}, where $K_{\mu}(x) = -i \frac{\delta D_{ov}(U_{\mu}^{(\alpha)})}{\delta \alpha(x)}|_{\alpha=0}$ and \\
 $U_{\mu}^{(\alpha)}(x) = e^{\alpha(x)} U_{\mu}(x)$.
\end{itemize}
These characteristics make overlap fermions particularly useful in lattice calculations of phenomena like nEDM and quark spin~\cite{Liang:2018pis,Liu:2021lke}, where maintaining chiral symmetry and accurately evaluating local topological charge are crucial.

\subsection{Nucleon EDM and cluster decomposition error reduction (CDER)}

The lattice $F_3$ form factors for the nEDM and pEDM are extracted from the three- and two-point correlation functions~\cite{Liang:2023jfj}
\begin{equation}
F_{3}(q^2) =\frac{2m}{E_f+m}\left\{\frac{2E_f}{q_i}\frac{{\rm Tr}
\left[\Gamma_{i}C_3(V_{\mu})^{Q}\right]}{{\rm Tr}\left[\Gamma_{e}C_2\right]}-\alpha_5\, G_{E}(q^2)\right\},
\end{equation}
where $G_E(q^2)$ is the electric form factor and $\alpha_5$ is the CP phase of the nucleon in Eq.~(\ref{alpha5}),
\begin{equation}   \label{GEangle5}
G_{E}(q^2)=\frac{2E_f}{E_f+m}\frac{{\rm Tr}
\left[\Gamma_{e}C_3(V_{\mu})\right]}{{\rm Tr}\left[\Gamma_{e}C_2\right]},\hspace{1cm} \alpha_5=\frac{{\rm Tr}\left[\gamma_{5}C_2^{Q}\right]}{2{\rm Tr}\left[\Gamma_{e}C_2\right]}.
\end{equation}
Here, $V_{\mu}$ is the EM current operator.
$C_3(V_{\mu})$ is the three-point correlator with nucleon interpolation fields at source and sink time positions of $0$ and $t_f$, and the vector current insertion occurs at time $t$. The source, sink and current have
momenta projections of $\vec{p}_i, \vec{p}_f$ and $\vec{q}$, respectively, such that the momentum transfer is $q^2$. $C_3(V_{\mu})^Q$ is the $C_3(V_{\mu})$ correlator with 
the topological charge $Q$ insertion. $C_2$ and $C_2^Q$ are the corresponding
two-point functions with the same source and sink at $0$ and $t_f$. 
$\Gamma_{e}=\frac{1+\gamma_{4}}{2}$ is the unpolarized spin projector, 
and $\Gamma_i={-i}\gamma_5\gamma_i\Gamma_{e}$ is the polarized projector along the $i$ direction. This formalism applies equally to both the neutron and the proton.

\begin{figure}
\begin{centering}
\includegraphics[scale=0.35]{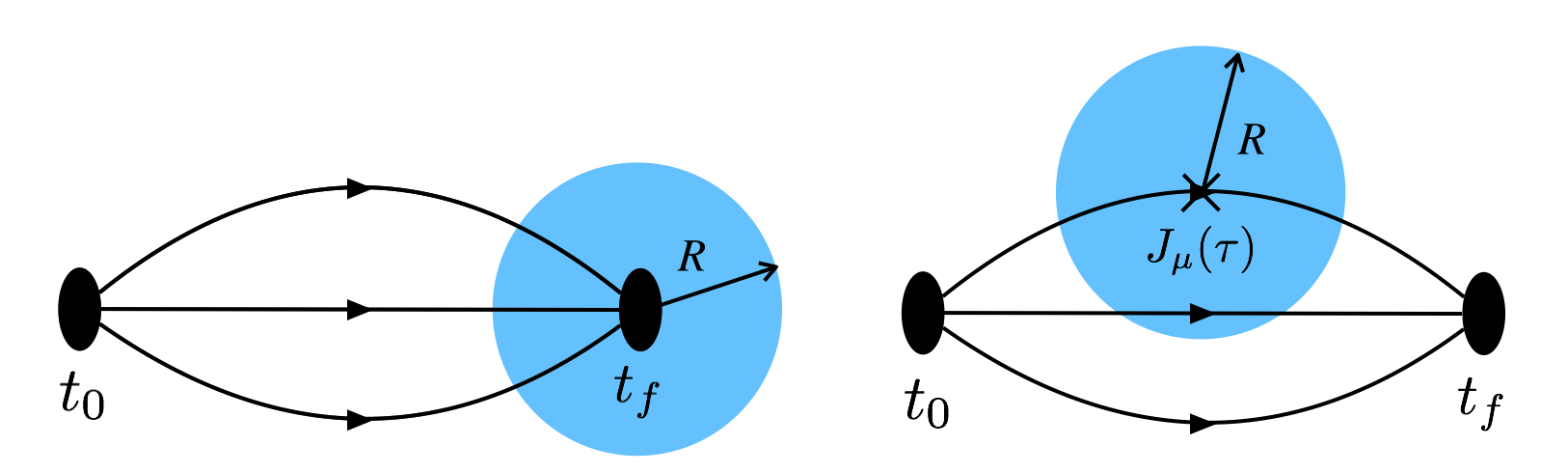}
\end{centering}
\centering{}
\vspace{0.5cm}
\caption{Illustration of the CDER technique used when computing the correlation functions with the local topological charge summed within a sphere with radius $R$. The left panel shows the calculation of $\alpha_5$ in $C_2^Q$, while the right panel depicts the CDER for $C_3(V_{\mu})^Q$ where $R$ is between the current and $q(x)$.
\label{fig:Illustration-of-CDER}}
\end{figure}

\begin{figure}[h]
\centering
\includegraphics[width=0.6\textwidth]{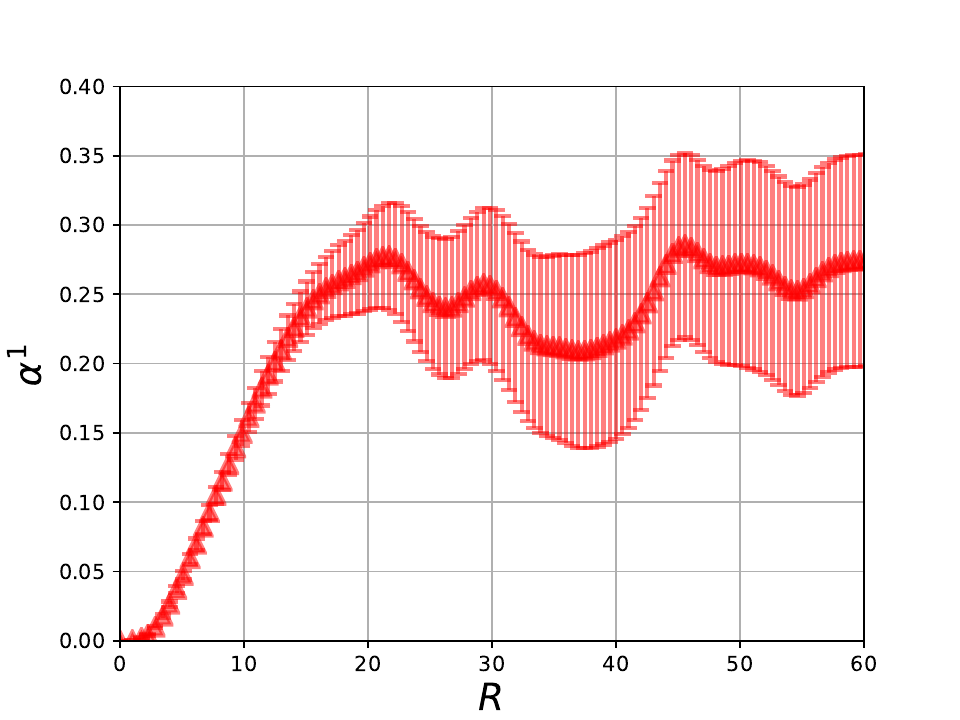}
\caption{The CP violation phase $\alpha_5$ calculated on the $48^3 \times 96$  DWF lattice as a function of cutoff $R$. For each $R$, the value is averaged from the source-sink time separation from 6 to 13.}
\label{fig_alpah5}
\end{figure}

One often invokes the locality argument to justify that
experiments conducted on Earth are not affected by events
on the Moon. This stems from the cluster decomposition principle (CDP), which states that if color-singlet operators
in a correlator are separated by a sufficient large space-like
distance, the correlator will be zero. In other words, the
operators become uncorrelated under these circumstances. To be
specific, it has been shown~\cite{Araki:1962zhd}  that under the assumptions of
translation invariance, stability of the vacuum, the existence
of a lowest non-zero mass and local commutativity, the correlator falls off exponentially with a correlation length of $1/M$ 
\begin{equation}   \label{CDP}
\begin{split}
&|\langle0|O_1(x_1)O_2(x_2)|0\rangle_s |\leq A r^{-\frac{2}{3}}e^{-Mr}
\end{split}
\end{equation}
for a large enough space-like distance $r=|x_1-x_2|$. In Euclidean space, $r$ can represent a space-time separation. 
On the other hand, a common problem arises in lattice QCD calculations of the hadron masses in annihilation
channels, where the signal diminishes over time while the noise remains constant. Additionally, the disconnected insertion (DI) calculation of the three-point function, such as the calculation of nEDM  with the $\theta$ term, encounters a noise issue due to the $\sqrt{V}$ fluctuation. These problems were identified~\cite{Liu:2017man} as stemming from the fact that the variance of the disconnected insertion has a leading intermediate vacuum state insertion, causing $O_1$ and $O_2$ in the disconnected insertion (DI) to fluctuate independently. This leads to a variance that is the product of their respective variances. In this case, the vacuum insertion is a constant, independent of $t$. This is why the noise remains constant over $t$ in DI. In the case of the correlator ${\rm Tr}(\gamma_{5}C_2^{Q})$, which is needed to obtain the CP phase $\alpha_5$ in Eq.~(\ref{GEangle5}), one can consider $Q$ as a sum of the local topological charge, i.e., $Q = \sum_x q(x)$, so that the correlator becomes
\begin{equation}
C_2^Q(R,\tau,t)=\langle \sum_{\vec{x}}\sum_{r<R}\mathcal{O}_N(\vec{x},t) q(\vec{x}+\vec{r'},\tau)\bar{\mathcal{O}}_N(\mathcal{G},0)\rangle.
\end{equation}
Here, $\bar{\mathcal{O}}_N(\mathcal{G},0)$ is the nucleon interpolation field for the grid source $\mathcal{G}$ at $t=0$. The variable
\mbox{$r = \sqrt{|\vec{r'}|^2 + (t-\tau)^2}$} represents the 4-D distance, and $r'$ is the spatial separation between the local charge $q(x)$ and the sink. R is the 4-D cutoff radius of the accumulated sum of $q(x)$. This is illustrated in Fig.~\ref{fig:Illustration-of-CDER}, where the left panel depicts the partial sum of $Q$ in the 4-sphere with radius $R$, while the right panel shows $R$ as the distance between the current position and the charge $q(x)$. The CP phase $\alpha_5$ on a $48^3 \times 96$ DWF lattice (with $a = 0.114\, {\rm fm}$, $m_{\pi} = 139\, {\rm MeV}$) using overlap fermions for the valence quarks and $q(x)$ from Eq.~(\ref{local_q}) is plotted as a function of $R$ in Fig.~\ref{fig_alpah5}. The resulting signal saturates after $R \sim 16$. Since the signal falls off exponentially due to the CDP, there is no further signal after this point, yet the error continues to accumulate. This suggests that the sum should be truncated at $R$. Cutting off the sum of $q(x)$ at this radius results in a factor of $\sim$ 3.6 times reduction in error compared to the case of reweighting
with the total topological charge as in Eq.~(\ref{GEangle5}). This approach can increase the
signal to noise ratio (SNR) by $\sqrt{V/V_R}$ where $V_R$ is the volume with the radius $R$. This variance reduction technique, known as the cluster decomposition error reduction (CDER), has been applied to the disconnected insertion (DI) calculations with quark loops~\cite{Liang:2018pis,Yang:2018nqn,Liang:2019xdx,Wang:2021vqy} and glue operators~\cite{Liang:2023jfj,Wang:2021vqy,Wang:2024lrm}. 


\begin{figure}[!htb]  
\centering
\includegraphics[width=1.0\textwidth]{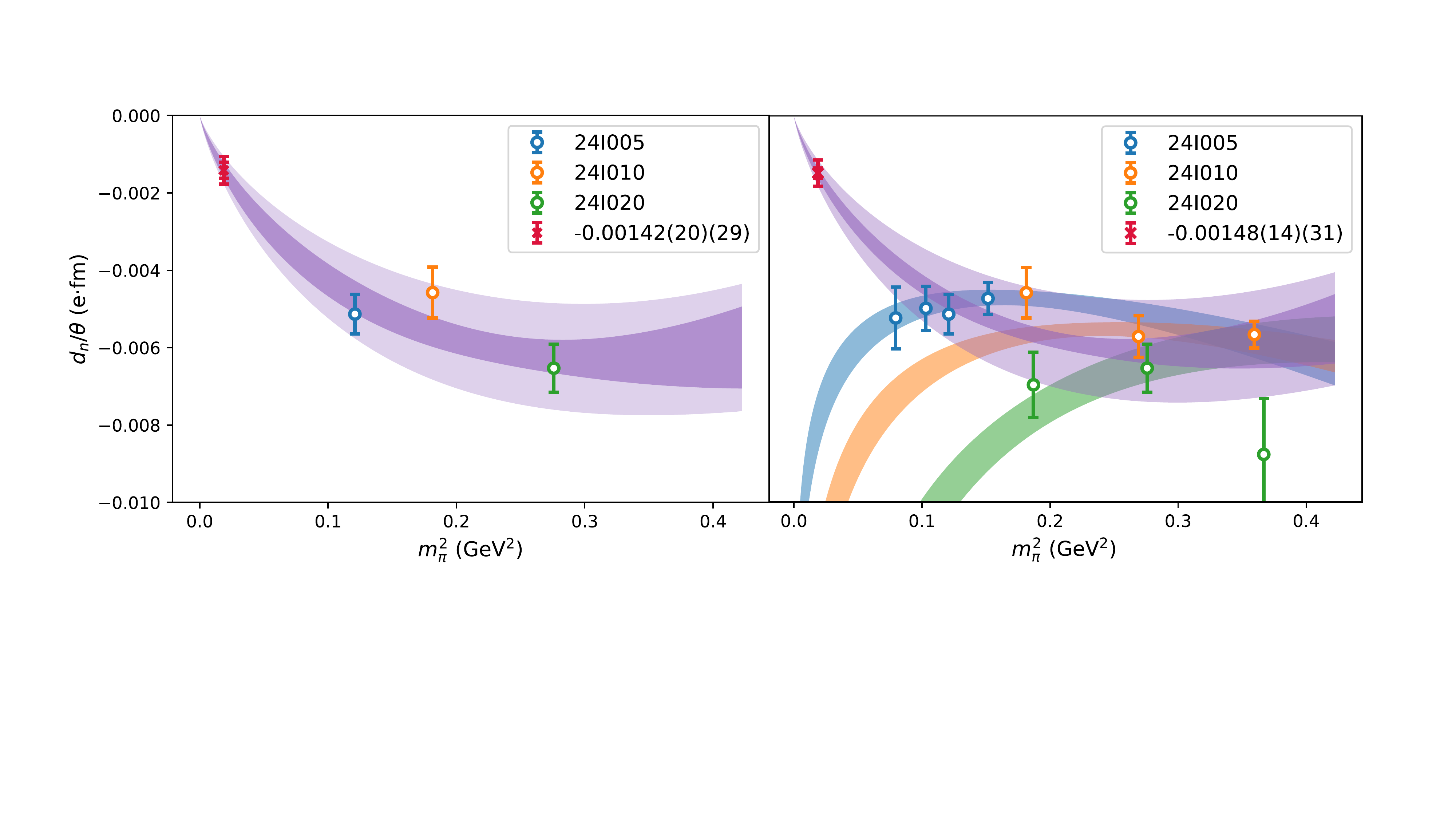}
\vspace*{-2cm}
 \caption{The neutron electric dipole moment as a function of $m_{\pi}^2$ is interpolated between the zero quark mass limit and three heavier pion masses. The left panel represents the unitary case, where the valence and sea quarks are the same. The right panel illustrates the case with 3 to 4 valence quarks for each sea quark ensemble in a partially quenched chiral perturbation interpolation. The red points indicate the values interpolated to the physical pion mass.
\label{nEDM}}
\end{figure}

Besides applying CDER to the calculation of nEDM and pEDM to reduce
errors from the DI of the topological charge~\cite{Liang:2023jfj}, additional variance reduction arises from the chiral nature of overlap fermions. As mentioned earlier, the anomalous Ward identity 
(Eq.~(\ref{AWI})) is preserved by the overlap fermion action and the $q(x)$ from the overlap Dirac operator. Consequently, the EDM at the chiral limit is zero according to Eq.~(\ref{theta-mass}) on the lattice at finite lattice spacing (This is not true with non-chiral fermions; in this case, one must first take the continuum limit to satisfy the AWI.) The results of the nEDM for three ensembles of the DFW configurations (see Table~\ref{tab1} for pion masses) are plotted in Fig.~\ref{nEDM} as a function of  $m_{\pi}^2$. The left panel shows the unitary case where the valence and sea pion masses are matched. Since $d_{n/p}$ anchored at the chiral limit is zero, a chiral interpolation can be performed to reach the physical pion point, incorporating a chiral log in the fitting. The result $d_n = - 0.00142(20)(29)\, \bar{\theta} \, e \cdot {\rm fm}$ is obtained, with the total systematic uncertainty arising from
excited-state contamination, the CDER cutoff, the $Q^2$ extrapolation and chiral interpolation. 

A multi-mass inversion algorithm for the overlap fermion admits the matrix inversion with multiple quark masses (as many as 18 has been tried~\cite{Dong:2001fm}), with an overhead of $\sim 8\%$ compared to the single inversion of the smallest mass~\cite{Ying:1996je}. Thus, it is advantageous to have several valence masses for one sea quark ensemble. 

Results from the partially quenched cases, with several valence quark masses for each sea mass, are plotted in the right panel of Fig.~\ref{nEDM}. It is observed that the chiral behaviors of the sea and valence pions seem to diverge, moving in opposite directions as the pion mass decreases. This behavior was predicted in a partially quenched $\chi$PT~\cite{OConnell:2005mfp} calculation, which indicates that the terms linear in $m_{\pi}^2$ have different signs for the sea and the valence pions. Consequently, the following form was used in the chiral interpolation, including both the valence and sea contributions.
\begin{equation}
d_n = c_1\, m_{\pi,s}^{2}\log\left(\frac{m_{\pi,v}^{2}}{m_{N}^{2}}\right)+c_{2}\,m_{\pi,s}^{2} 
+  c_{3}\left(m_{\pi,v}^{2}-m_{\pi,s}^{2}\right),
\end{equation}
where $m_{\pi,v}$ and $m_{\pi,s}$ are the valence and sea pion masses, respectively. With additional partially quenched data, the nEDM is fond to be $d_n = - 0.00148(14)(31)\, \bar{\theta} \, e \cdot {\rm fm}$, reflecting a 30\% reduction in the statistical error compared to the unitary case. The 10-sigma result (statistical error) represents a significant improvement over the previous 2-sigma result~\cite{Draper:2006wb}. This enhancement is primarily due to the combination of CDER, chiral fermion interpolation, and the inclusion of partially quenched data.  The proton EDM is also obtained in the same calculation, yielding $d_p =0.038(11)(8)\, \bar{\theta} \, e \cdot {\rm fm}$.

A precise determination of $d_n$ and $d_p$ is important for systems that are sensitive to the iso-singlet combination  $d_n + d_p$, where the one-loop chiral perturbation contributions almost cancel. Since the disconnected insertion (DI) of the EM current is small due to the cancellation of the charges of the $u,d$ and $s$ quarks, the connected insertion (CI) calculation should provide a good estimate. Experimental efforts
have been initiated in the Storage Ring COSY at the Forschungszentrum in 
J\"{u}lich~\cite{CPEDM:2019nwp} to measure of EDMs of charged particles, such as the proton, deuteron and possibly ${}^3{\rm He}$. This supports the notion that it is necessary to  measure EDMs in various systems experimentally, as they might be sensitive to different sources of CP violation.

\subsection{Chiral perturbation theory and CP-violating $\pi {\rm NN}$ coupling}

\begin{figure}[!htb]  
\centering
\includegraphics[width=0.9\textwidth]{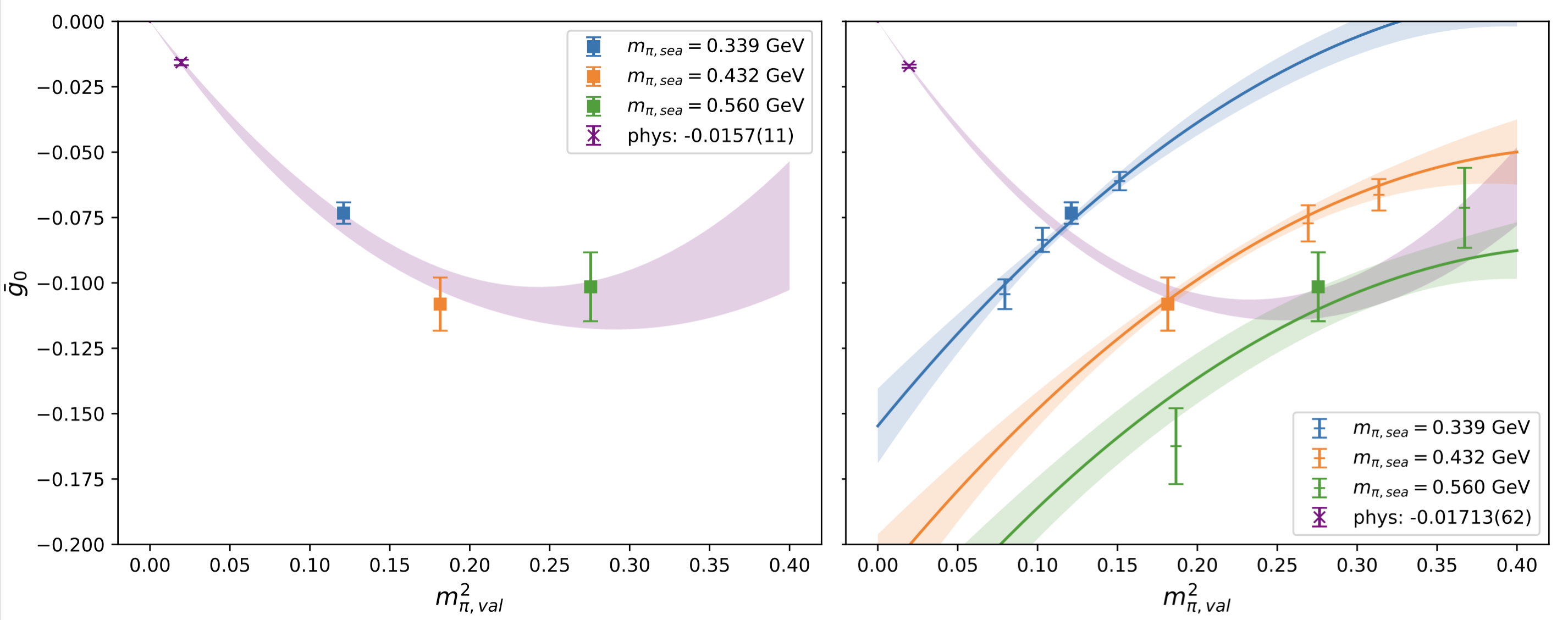}
 \caption{The same as in Fig.~\ref{nEDM} for the CP-violating $\pi {\rm NN}$ coupling from the $\theta$ term. The left panel shows the  unitary case, while the right panel presents the partially quenched results.
\label{fig:g_0}}
\end{figure}

The extraction of solid information on possible new sources of CP violation from EDM
measurements involves dynamics across a wide variety of scales, from the new physics scale $\Lambda_{\cancel{CP}}$ to the electroweak (EW) and QCD scales, down to the atomic scale. At the QCD scale, the interpretation of EDM experiments involving more than one nucleon additionally depends on CP-violating (CPV) nucleon-nucleon interactions. Heavy Baryon Chiral Perturbation Theory (HB$\chi$PT)~\cite{Jenkins:1990jv} has been extended to include  parity and time reversal breaking Lagrangians. The chiral power counting predicts that these interactions mainly depend on one-pion-exchange contributions involving CP violating (CPV) pion-nucleon vertices~\cite{Mereghetti:2010tp, Maekawa:2011vs,deVries:2012ab,Bsaisou:2014oka}. 
 Calculations of the CPV pion-nucleon coupling constants are therefore as important as those of nucleon EDMs. The relevant CPV pion-nucleon coupling Lagrangian includes the following terms
\begin{equation}
\mathcal{L}_{\rm eff}^{\cancel{CP}} = \bar{g}_0 \overline{N} \boldsymbol{\tau \cdot \pi} N
+ \bar{g}_1  \pi_0 N + \bar{g}_2  \pi_0 \tau^3 N,
\end{equation}
where the low-energy constants (LECs) $\bar{g}_0, \bar{g}_1$ and $\bar{g}_2$ represent the iso-scalar, iso-vector, and iso-tensor CP-odd pion-nucleon couplings, respectively.  
The first estimate of nEDM was based on the pion loop of the nucleon with the CP-violating $\pi {\rm NN}$ coupling $\bar{g}_0$ from the $\Xi$ and nucleon mass difference~\cite{Crewther:1979pi} and from mixing with the $\Delta$ resonance~\cite{Baluni:1978rf}. These LECs can be obtained from baryon mass differences by comparing the $\bar{g}_0$ term with the expansion of the quark mass term, which is modified by the presence of the $\bar{\theta}$ term in 
Eq.~(\ref{theta-mass}). The mass and CPV terms appear in the HB$\chi$PT Lagrangian only at $O(p^2)$, and if the decuplet terms are neglected, the expansion of $\bar{\theta}$ in the modified mass $\chi = 2B(M + i m^*\, \bar{\theta})$ in the mass coupling term gives
\begin{eqnarray}    \label{M_N}
\bar{g}_0 &=& - \frac{(\Delta M_N)_{\rm QCD}}{2 F_{\pi}} 
\frac{m^*\,\bar{\theta}}{\overline{m}\epsilon}  \nonumber \\
   &=& - \frac{(\Delta M_N)_{\rm QCD}}{2 F_{\pi}} 
 \left(\frac{1 - \epsilon^2}{2\epsilon}\right) \bar{\theta} + O(\frac{\overline{m}}{m_s})
\end{eqnarray}
where $\overline{m} = \frac{m_u + m_d}{2}$, $\epsilon = \frac{m_d-m_u}{m_d+m_u}$ and $(\Delta M_N)_{\rm QCD}$ is the neutron and proton mass difference in QCD, accounting for isospin breaking due to the different $u$ and $d$ quark masses without QED effects. While this leading-order result in Eq.~(\ref{M_N}) has been known for some time~\cite{Crewther:1979pi,Engel:2013lsa,deVries:2012ab}, it has been observed that considering the isospin splitting in $SU(3)$ $\chi$PT to the next-to-next-leading order (${\rm N^2LO}$), i.e., $\mathcal{O}(m_q^2)$, the relation between $\Delta M_N$ and $\bar{g}_0$ holds to within a few percent~\cite{deVries:2015una}.
Using lattice input for $\Delta M_N$, $m_u$, and $m_d$~\cite{BMW:2014pzb,Brantley:2016our}, a precisely determined $\bar{g}_0$ is obtained~\cite{deVries:2015una,Brantley:2016our}
\begin{equation}  \label{g_0}
    \bar{g}_0 = -15.5\, (2.0) \, (1.6)\times 10^{-3}\, \bar{\theta}.
\end{equation}
This CPV $\pi {\rm NN}$ coupling can be obtained from the Schiff moments of nEDM and pEDM. It can also be determined from the pseudoscalar form factor
 in the nucleon in the presence of the $\theta$ term. Assuming pion dominance of the pseudoscalar form factor, $\bar{g}_0$ can be obtained from the matrix element
 \begin{equation}
     \langle N (p')|\bar{\psi}\,i \gamma_5 \psi (-i \bar{\theta}\, Q) |N(p) \rangle = \bar{u}(p')\, \bar{g}_P (q^2)\, u(p),
\end{equation}
where $q^2$ is the momentum transfer. Assuming pion pole dominance, we have 
\begin{equation}
    \bar{g}_P (q^2) = \left(\frac{m_{\pi}^2 f_{\pi}}{2 m_q}\right)\frac{\bar{g}_0}{- q^2 + m_{\pi}^2}.
\end{equation}
 A recent lattice calculation of the pseudoscalar form factor with the the topological charge insertion has been carried out~\cite{Liang:20241003} by the $\chi$QCD collaboration based on the same ensemble of lattices used for the nEDM and pEDM calculation~\cite{Liang:2023jfj}. They obtained $\bar{g}_0 = 
 - 0.01713 (62)\, \bar{\theta}$,
 which is consistent with the value  in Eq.~(\ref{g_0}), and the statistic error is about 3 times smaller.
 Similar to Fig.~\ref{nEDM}, the pion mass dependence of $\bar{g}_0$ on the sea and valence pion masses is plotted in Fig.~\ref{fig:g_0}.

\section{Weinberg operator}

After integrating out heavy quarks and Higgs bosons, a dimension-6 CPV term emerges, specifically Weinberg's three-gluon operator~\cite{Weinberg:1989dx}.
\begin{equation}   \label{W-operator}
    O_W = k_W{\rm Tr}\, (\lbrack G_{\mu\rho}, G_{\nu\rho}\rbrack \tilde{G}_{\mu\nu}).
\end{equation}
The Weinberg operator can potentially contribute significantly to nEDM and pEDM since it is a pure glue operator. As such, it is not suppressed by small quark masses, unlike the $\theta$ term. 

Calculating operators on the lattice requires renormalization, and dimension-6 operators can mix with lower dimensions operators, leading to power divergences in $1/a$. The standard non-perturbative lattice renormalization scheme is the regularization-independent momentum-subtraction scheme (RI/MOM)~\cite{Martinelli:1994ty}. In this scheme, the renormalization conditions are imposed on quark and gluon correlation functions, computed in a fixed gauge, with off-shell external states of large space-like virtualities. 

Given that it is an off-shell scheme with gauge fixing, the renormalization of the Weinberg's operator is complicated~\cite{Cirigliano:2020msr}. It mixes with three dimension-4 operators, and for $N_f \ge 3$, one dimension-5 operator. Additionally, it also mixes with thirty dimension-6 operators, twenty of which are `nuisance' operators -- some of which vanish with the equation of motion (EOM), while some others are gauge non-invariant. Although  mixing with gauge-variant operators can be avoided by employing the background-field method~\cite{Abbott:1980hw},
the remaining mixing patterns still present significant challenges for the numerical aspects of the lattice calculation. 

\subsection{Gradient flow}

An alternative to the RI/MOM scheme is gradient flow~\cite{Narayanan:2006rf,Luscher:2010iy,Luscher:2013cpa}. This is a gauge-invariant renormalized smoothing procedure that modifies the fields at short distances through a diffusion equation. The evolution of the gauge field is given by:
\begin{equation}
    \partial_t B_{\mu} = D_{\nu} G_{\nu\mu}, \hspace{2.8cm} B_{\mu}|_{t =0} = A_{\mu}, 
\end{equation}
where,
\begin{equation}
    G_{\mu\nu} = \partial_{\mu}B_{\nu} - \partial_{\nu}B_{\mu} + [B_{\mu}, B_{\nu}],      \hspace{1cm} D_{\mu} = \partial_{\mu} + [B_{\mu}, \cdot],
    \end{equation}
The gradient flow equation introduces a fifth dimension $t$ (or $t_{\rm gf}$), which is distinct from the Euclidean time. For the fermions, the flow equations are
\begin{eqnarray}
    \partial_t \chi &=& D_{\mu}D_{\mu}\chi, \hspace{2.5cm} \chi|_{t=0} = \psi, \nonumber \\
    \partial_t \overline{\chi} &=&  \overline{\chi}\, \overleftarrow 
    {D}_{\mu}\overleftarrow{D}_{\mu} \hspace{2.5cm} \overline{\chi}|_{t=0} = \overline{\psi}.
\end{eqnarray}
The initial values of the flowed fields $B_{\mu}$ and $\chi$ are $A_{\mu}$ and
$\psi$, respectively, which correspond to those in the un-smoothed 4-dimensional space. The flow time $t$ has the dimension of length squared. Besides being a gauge-invariant procedure, it provides a
quantum definition of the topological charge $Q$, which takes on an integer value for each gauge configuration after a certain flow time, i.e., $\partial_t Q(t) = 0$~\cite{Ce:2015qha}. Furthermore, it is a powerful tool for defining the multiplicatively renormalized higher dimensional operators without mixing with lower-dimensional operators~\cite{Luscher:2013vga}. 

While the gauge field $B_{\mu}$ at $t > 0$ does not require renormalization, the fermion fields do~\cite{Luscher:2013cpa}. The divergent part of the fermion field renormalization factor $Z_{\chi}$ has been calculated at one loop~\cite{Luscher:2013cpa,Makino:2014taa,Monahan:2017hpu} and at two loops~\cite{Artz:2019bpr}, while the finite piece has been computed in \cite{Rizik:2020naq}. This suggests that the flowed fermion fields do not exhibit power diverges. To explore where the power divergences have gone, it is useful to consider the short flow-time behavior of the composite operators. 

The short flow-time expansion (SFTE)~\cite{Luscher:2013vga} of the gauge-invariant renormalized local operators $(O_i)_R (t)$ can be expanded as
\begin{equation}   \label{SFTE}
    (O_i)_R (t)\,\, \stackrel{t\rightarrow0}{\sim}\,\, \sum_i c_{ij}(t) (O_j)_R (0),
\end{equation}
where $(O_j)_R (0)$ are renormalized operators at $t=0$ that have the same symmetry as  $(O_j)$ at $t$ and dimensions that are equal to or lower than that of $(O_i) (t)$. The SFTE is an operator product expansion around $t =0$, with Wilson coefficients $c_{ij}(t)$ that can be computed perturbatively. The flow-time dependence is encapsulated in these Wilson coefficients~\cite{Luscher:2013vga}. 

In lattice perturbation theory, there is only one scale --- lattice spacing $a$ --- which serves the roles of both the cutoff and the renormalization scale. It is non-trivial to disentangle these roles, especially when the operator is mixed with lower dimensional ones, leading to power divergence in $1/a$. Gradient flow provides a solution by introducing a new scale in the flow time $t$, distinct from the lattice cutoff scale. In the continuum limit, all correlators are functions of the flow time with a scale 
$\mu^2 \propto 1/t$. The SFTE offers a method to extract renormalized operators at $t=0$ and their mixings from operators calculated at $t>0$ which are finite. The challenge associated with the renormalization of operators at finite $a$ in other renormalization scheme is replaced by the determination of the Wilson coefficients $c_{ij}(t)$ in 
Eq.~(\ref{SFTE}). One desirable feature here is that the analysis can be performed in the  continuum, avoiding cumbersome chiral symmetry breaking effects associated with non-chiral fermion actions at finite $a$. 

The perturbative SFTE for the Weinberg operator has been calculated~\cite{Rizik:2020naq} for the mixing of 
lower-dimensional operator up to  $\mathcal{O}(g^2)$. To this order, only the topological charge term contributes
\begin{equation}  \label{O_W}
    \mathcal{O}_{W}^R(t)\stackrel{t\rightarrow0}{\sim} -
\frac{45}{8} g^2 \frac{k_W}{k_q}\frac{C_2(A)}{(4\pi)^2}\left\{\frac{1}{t}+
\frac{8}{15}p^2\left[\log(2p^2t)+\gamma_E-\frac{35}{16}\right]\right\}
\mathcal{O}_{q}^R(0)+\cdots.
\end{equation}
where $k_q$ is defined in the topological charge operator $O_{q} = k_q {\rm Tr} \lbrack G_{\mu\nu} \tilde{G}_{\mu\nu}\rbrack$ and $C_2 (A) = N$ for $SU(N)$.  

Notably, there is a $1/t$ term, indicating that the Wilson coefficient diverges as $t \rightarrow 0$. This contrasts with the $1/a^2$ power divergence found in off-shell renormalization schemes, such as RI/MOM. The susceptibility of the Weinberg operator in the gradient flow exhibits a divergent behavior as $t$ becomes small, in contrast to the flat behavior of topological susceptibility as a function of the flow time~\cite{Dragos:2017wms}. This aligns with the expected $1/t$ behavior of $\mathcal{O}_{W}^R(t)$ in 
Eq.~(\ref{O_W}). 

\begin{figure}[bt]  
 \vspace*{-2cm}
  \begin{center}
    \includegraphics[width=0.9\hsize]{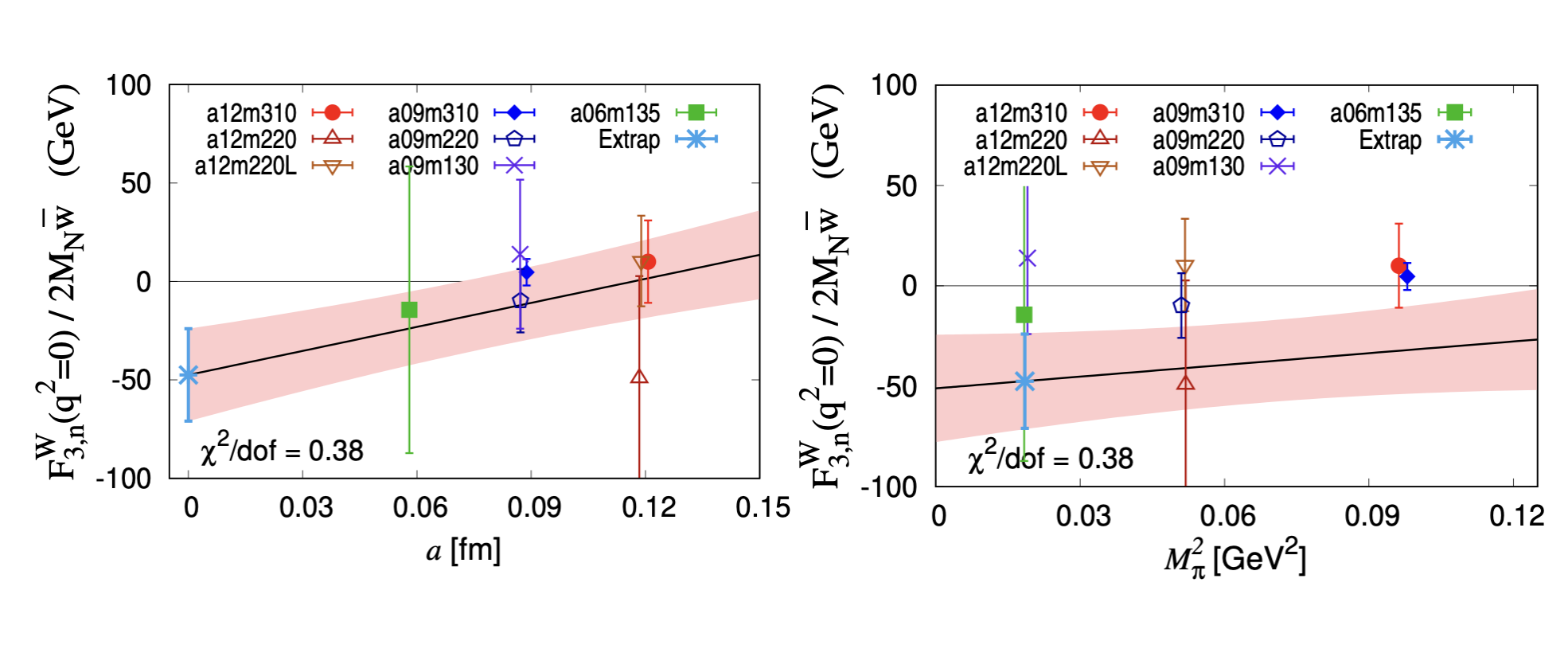}
  \end{center}
  \caption{The simultaneous continuum (left) and chiral (right) extrapolation of nEDM due to the Weinberg operator in the gradient-flow scheme at $\tau_{\rm gf}\approx 0.34$ fm. The $\overline{w}$ is $k_W$ in Eq.~(\ref{W-operator}).
  \label{fig:Weinberg}}
\end{figure}

A preliminary lattice gradient flow calculation of $F_3$ using the Weinberg operator in place of the $\theta$ term has been attempted~\cite{Bhattacharya:2022whc}. The calculations were conducted with clover valence fermions on multiple HISQ lattices with different lattice spacings and quark masses. The results are extrapolated to the continuum limit and are
plotted in Fig.~\ref{fig:Weinberg}. This represents the raw lattice data at a fixed gradient flow time \mbox{$t_{\rm gf} \approx 0.34$ fm.}  

Completing the calculation is a challenging task, as it requires
consideration of mixing with lower-dimensional operators exhibiting $1/t$ behavior, operators of same dimension with logarithmic $t$ dependence, and higher dimension operators with $\mathcal{O}(t)$ dependence. Furthermore, these mixed operators need to be renormalized at $t =0$.  

\section{Quark electric dipole moment and quark chromoelectric dipole moment}

    At dimension 5, there are two CPV operators
\begin{equation}
    \mathcal{L}_q^{\cancel{CP}} = - \frac{i}{2} \sum_q d_q\, \bar{q}\sigma_{\mu\nu} \tilde{F}_{\mu\nu}q - \frac{i}{2} \sum_q \bar{d}_q\, \bar{q}\sigma_{\mu\nu} \tilde{G}_{\mu\nu}q,
\end{equation}
where the first term represents the quark EDM (qEDM) with $\tilde{F}_{\mu\nu}$ being the dual of the electromagnetic field tensor $F_{\mu\nu}$ (note that $\sigma_{\mu\nu} \tilde{F}_{\mu\nu}= -\sigma_{\mu\nu} \gamma_5 F_{\mu\nu}$.) This entails the evaluation of the tensor charge of the nucleon, a task well-suited for lattice calculations. The tensor charge of the nucleon can be obtained from the nucleon matrix element of the tensor current 
\begin{equation}
    \langle N(p)| \bar{q}\, \sigma_{\mu\nu} q|N(p)\rangle = g_T^q\, \bar{u}(p)\,  \sigma_{\mu\nu} u(p).  
\end{equation}
Here, $g_T^q$ is the tensor charge for the quark flavor $q$. It represents the first Mellin moment of the collinear transversity parton distribution function (PDF) and can be extracted from the transversity parton distribution from deep inelastic scattering.

The first lattice calculation of the $g_T^q$ for the $u,d$ and $s$ quarks, including all systematics, was performed by the PNDEME collaboration~\cite{Bhattacharya:2015esa}. These calculations utilized valence clover fermions on $N_f = 2+1+1$ HISQ configurations. More precise results were updated~\cite{Gupta:2018lvp} using 11 ensembles of configurations for the connected insertion and 7(6) ensembles for the strange (light) quarks in the disconnected insertion. A simultaneous fit in terms of the lattice spacing and the light-quark masses at the physical pion mass of 135 MeV, in the $\overline{\rm MS}$ scheme at 2 GeV, gives~\cite{Gupta:2018lvp,FlavourLatticeAveragingGroupFLAG:2021npn}
\begin{equation}
    g_T^u = 0.784(28)(10), \hspace{1cm} g_T^d =  -0.204(11)(10), \hspace{1cm} g_T^s = -0.0027(16)
\end{equation}

The quark chromoelectric dipole moment (qcEDM) exhibits a similarly complicated renormalization pattern~\cite{Bhattacharya:2015rsa} as the Weinberg term. The SFTE in Eq.~(\ref{SFTE}) for the qcEDM operator is given by:
\begin{equation}
(O_{\rm qCEDM})_R (t)\,\, \stackrel{t\rightarrow0}{\sim}\,\, c_{CP} P_R(0) + c_{Cq} (O_q)_R (0) + \cdots,
\end{equation}
which indicates mixing with the lower dimensional operators 
$P = k_P \overline{\psi} \gamma_5 \psi$ and 
\mbox{$O_q = (k_q/4)\, \epsilon_{\mu\nu\alpha\beta}  G_{\mu\nu}^a G_{\alpha\beta}^a$}. The ellipsis ($\cdots$) denotes mixing with same dimension 5 and higher dimensional operators. 

The leading order in a perturbative calculation~\cite{Rizik:2020naq} yields:
\begin{equation}
	\begin{aligned}
		\mathcal{O}_C^R(t)\stackrel{t\rightarrow 0}{\sim} \,\,&6ig^2\frac{k_C}{k_P}\frac{C_2(F)}{(4\pi)^2}\left\{\frac{1}{t}+p^2\left[\log\left(2p^2t\right)+\gamma_E-\frac{11}{4}\right]\right\}\mathcal{O}_P^R(0)\\
		&+4ig^2\frac{k_C}{k_q}\frac{m}{(4\pi)^2}\left[\log(2p^2t)+\gamma_E-1\right]\mathcal{O}_q^R(0)+ \cdots
	\end{aligned}
\end{equation}
   The Wilson coefficient $c_{CP}$ confirms the expectation that the leading small $t$ contribution
   of $\mathcal{O}_C^R(t)$ is given by the pseudoscalar density, with a leading $1/t$ divergence.

\begin{figure}[tb]  
 \vspace*{-2cm}
  \begin{center}
    \includegraphics[width=0.9\hsize]{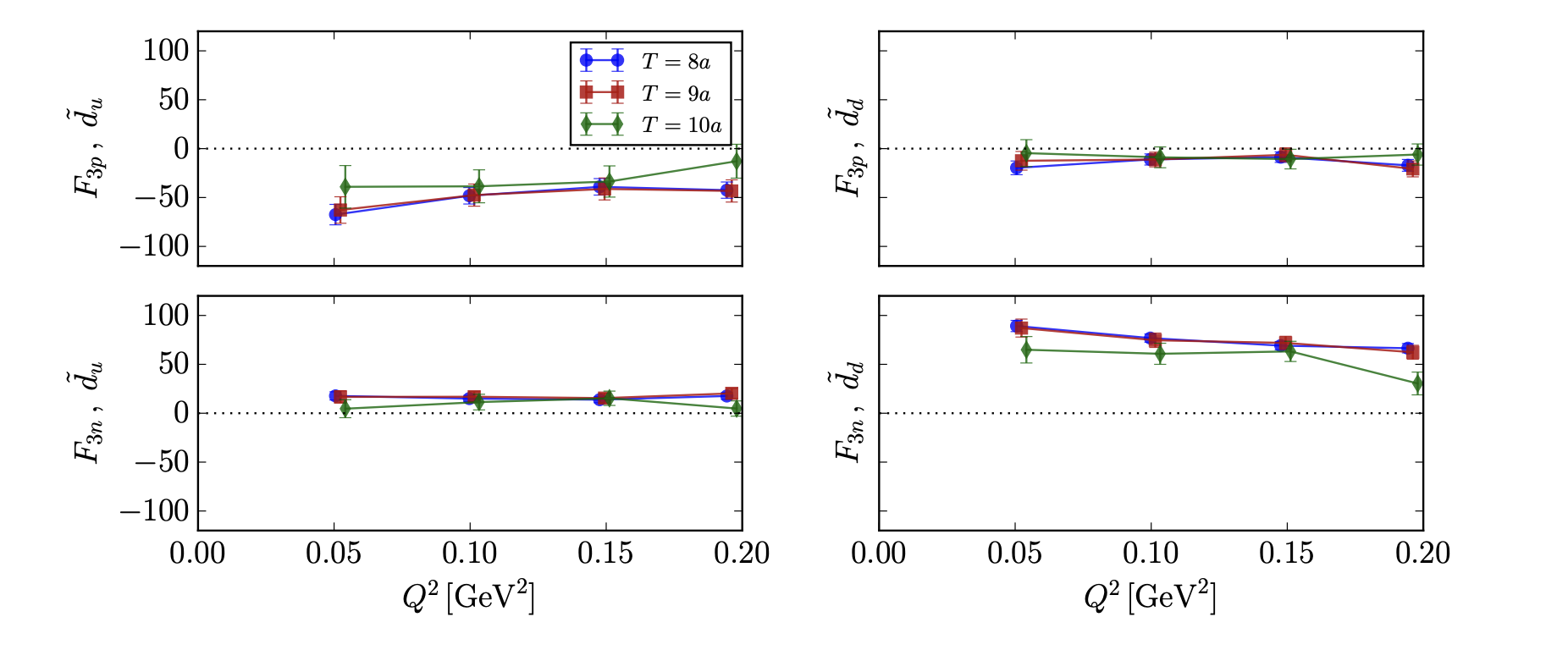}
  \end{center}
  \caption{The contributions of the $u$ and $d$ quark to the bare $F_3$ form factors as functions of $Q^2$ for the neutron ($n$) and proton ($p$) are shown with different source-sink time separations $T$. The left panels display the contributions from the $u$ quarks, while the right panels display those from the $d$ quarks. This is a direct four-point function calculation with the qcEDM operator insertion~\cite{Syritsyn:2019vvt}.
  \label{fig:qcEDM-BNL}}
\end{figure}

   There have been preliminary attempts to calculate qcEDM on the lattice\cite{Bhattacharya:2015ftz,Abramczyk:2017oxr,Syritsyn:2019vvt,Izubuchi:2020ngl,Bhattacharya:2023qwf}. One such calculation was performed by the Stony Brook-BNL group~\cite{Syritsyn:2019vvt} using a DWF lattices with a lattice spacing of $a=0.1141(3)$ fm at the physical pion mass. This involved a direct four-point function calculation with the qcEDM operator insertion. Only the connected insertions were calculated for both the CP-odd and CP-even correlators.  Fig.~\ref{fig:qcEDM-BNL} presents the contributions of the $u$ and $d$ quark to the bare $F_3$ form factors for the neutron and proton at different source-sink time separation $T$. A crucial future plan for this approach will necessarily involve renormalization and mixing with operators of the same and lower dimensions in the RI-SMOM scheme~\cite{Bhattacharya:2015esa}.

Another recent qcEDM calculation was carried out by the Los Alamos group~\cite{Bhattacharya:2023qwf}. The calculation utilized  clover fermions on four 2+1+1-flavor HISQ ensembles. Three of these ensembles have lattice spacings of $a = 0.12, 0.09$ and 0.06 fm, with a   pion masses around 300 MeV. The fourth ensemble, also at $a= 0.12$ fm, has a pion mass at 227 MeV. These configurations were employed for continuum extrapolation and the study of poin mass dependence. They developed a Schwinger source method~\cite{Bhattacharya:2015ftz,Gupta:2017anz} that incorporates the qcEDM operator, a quark bilinear, into the action to calculate the nucleon EDM. This involves reweighting the ratio of the determinants between the modified and the original ones. Additionally, the three-point function for the vector current must be calculated using a combination of quark propagators --- with and without the qcEDM operator in the Dirac matrix --- to examine the linear region of the small scale parameter multiplied by the qcEDM operator. Since this approach uses a mixed action with non-chiral valence fermions and the operators are not $O(a)$ improved, there is an $O(a)$ error~\cite{Chen:2007ug}. 
The axial Ward identity for the Wilson-clover action includes an additive term coresponding to the dimension-5 qcEDM operator~\cite{Karsten:1980wd,Bochicchio:1985xa,Guadagnoli:2002nm}. Utilizing this identity, the isovector nEDM from the qcEDM is obtained for the power-divergence subtracted $F_3$ form factor, which is proportional to that from the pseudoscalar operator. However, it is found that the omitted $O(a^2)$ error can be as large as $\sim 25\%$~\cite{Bhattacharya:2023qwf}. 

\begin{figure}[bt]  \label{qCEDM-LANL}
  \begin{center}
    \includegraphics[width=0.9\hsize]{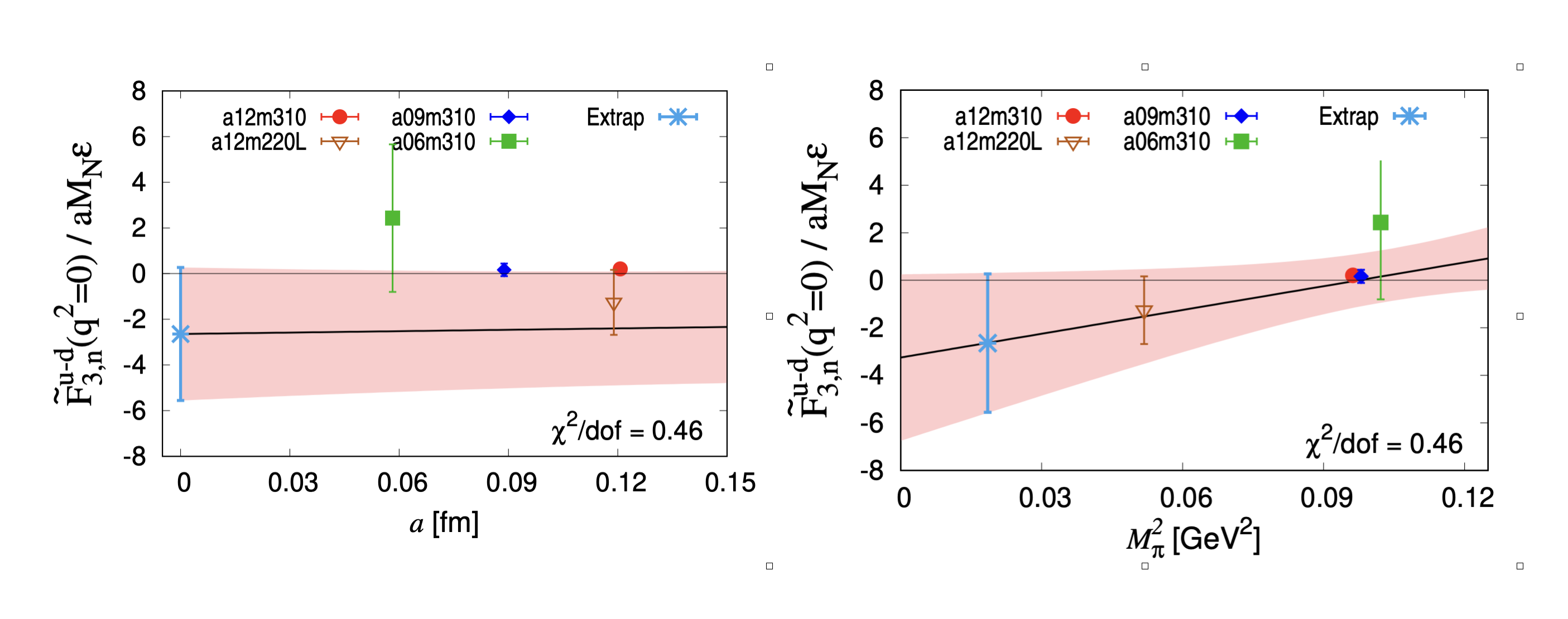}
  \end{center}
  \caption{The simultaneous continuum (left) and chiral (right) extrapolations of the isovector nEDM, resulting from the qcEDM action, are calculated using the Schwinger source method~\cite{Bhattacharya:2023qwf}.
  \label{fig:qCEDM-LANL}}
\end{figure}

The results for the isovector neutron $F_3$ form factor at $q^2=0$ are shown in Fig.~\ref{fig:qCEDM-LANL}, with the continuum extrapolation presented in the left panel and $m_{\pi}^2$ dependence in the right panel.
They obtained nEDM $F_3(0)/2M_N\epsilon = -2.6(2.9)$, where $\epsilon$ is the dimensionful parameter which multiplies the qcEDM term. The study also considered a fit accounting for the $\pi N$ excited state contamination, which resulted in a 5-fold increase in $F_3(0)/2M_N\epsilon$. 

The isovector case considered in this work allowed several simplifications: the determinant reweighting and the mixing with the topological term, and other dimension-5 operators are avoided. In the future when individual $u,d,s$ quarks are considered, these complications need to be addressed~\cite{Bhattacharya:2015ftz}.

\section{Summary Points}
\begin{enumerate}
\item Any experimental observation of a permanent EDM in non-degenerate systems, such as hadrons, would signal CP violation and indicate new physics beyond the 
Standard Model. This will have implications for the origin of baryon-antibaryon asymmetry in the Universe. Observations in complimentary systems will be necessry to identify CP violations at low energy and systematically connect to the underline BSM models . In the next decade, several orders of magnitude in sensitivity could be
achieved over current bounds for nucleons, electrons, atoms, 
and molecules~\cite{Alarcon:2022ero}. Theoretical efforts in
Standard Model Effective Theory (SMEFT), lattice QCD, low-energy effective theories (LEFT), and nuclear many-body calculations will be integral to this endeavour.

\item Lattice QCD has matured to the point where it can calculate nucleon matrix elements with controlled systematic errors. We have observed significant improvements in the calculations of nEDM and pEDM with the $\theta$ term, despite the smallness of the signals. An important factor is the adoption of chiral fermions, which ensures that nEDM and pEDM are zero in the chiral limit at finite $a$. Additionally, variance reduction with the CEDR technique can decrease the variance of the signal by a volume factor. The agreement of the CP-violating $\pi {\rm NN}$ coupling from lattice calculations  with predictions from the baryon spectrum serves as a consistency check between chiral perturbation theory and lattice calculations.

\item Lattice calculations of the Weinberg term and the quark chromo-electric dipole moment (qcEDM) have been undertaken. A major challenge for these operators is the complicated renormalization pattern. As higher dimensional operators, they mix with both other operators of the same dimension and lower-dimensional ones, leading to divergences in powers of $1/a$ in off-shell renormalization schemes, such as
RI/MOM. The gradient flow with short flow-time expansion (SFTE) is gauge invariant and makes the renormalization of higher-dimensional operators more manageable. The power divergence of $1/a^2$ is reflected in the $1/t$ behavior in the SFTE as $t$ approaches zero. In contrast to the small signal of nEDM due to the $\theta$ term, which is
$\sim 10^{-3}$, the nEDM from the Weinberg term and qcEDM appear to be on the order of $O(1)$ to $O(10)$. This is encouraging news,indicating that the numerical efforts required to calculate these terms are less demanding.

\end{enumerate}

\section{Future Issues}
\begin{enumerate}
\item Despite the statistical error of the current calculation of nEDM from the $\theta$ term is at the 10\% level, it is based on chiral interpolation from heavier pion masses greater than 300 MeV. The calculation should be performed for ensembles at the physical pion mass, including systematic errors from continuum and infinite volume extrapolations.

\item The non-perturbative approach to the gradient flow has been developed~\cite{Hasenfratz:2022wll}. It would be desirable to tackle the Weinberg, qcEDM, and four-fermion interactions with the gradient flow - SFTE scheme for renormalization and mixing.

\item Chiral fermions have the desirable feature of satisfying current algebra. We have seen their advantage in calculating nEDM with the $\theta$ term. They simplify renormalization, and their $O(a^2)$ error is small. In particular, the $O(a^2)$ of the overlap fermion is the smallest compared to other fermions in spectroscopy~\cite{Draper:2006wb}  and in the continuum extrapolation of the muon $g$-2 calculation~\cite{Wang:2022lkq}. It would be worthwhile to consider using chiral fermions for nEDM calculations involving higher-dimensional operators. For the lattice results to be fully credible and trustworthy, it would be essential to have multiple fermion actions and renormalization schemes to 
cross-check the final results at the continuum and infinite volume limits. 

\end{enumerate}

\section*{DISCLOSURE STATEMENT}
The authors are not aware of any affiliations, memberships, funding, or financial holdings that
might be perceived as affecting the objectivity of this review. 

\section*{ACKNOWLEDGMENTS}
The author thanks T. Drape for reviewing the manuscript and J. Liang of $\chi$QCD collaboration for providing the preliminary results on the CP-violating $\pi {\rm NN}$ coupling before publication. He is grateful for discussions with K. Fuyuto, F. He, 
E. Mereghetti, A. Shindler, S. Syritsyn, and \mbox{A. Walker-Loud.} The author also thanks R. Gupta, T. Bhattacharya, and
S. Syritsyn for their permission to reproduce their figures in Figs.~\ref{fig:Weinberg}, \ref{fig:qcEDM-BNL} and \ref{fig:qCEDM-LANL}.  

This work is supported in part by
the U.S. Department of Energy, Office of Science, Office of Nuclear Physics, under Grant No. DE-SC0013065.
The author also acknowledges partial support by the U.S.
Department of Energy, Office of Science, Office of Nuclear Physics under the umbrella of the Quark-Gluon
Tomography (QGT) Topical Collaboration, with Award
No. DE-SC0023646.

%


\bibliography{Review_nEDM_arXiv}

\end{document}